# Enhanced Piezoelectric Response in Hybrid Lead Halide Perovskite Thin Films via Interfacing with Ferroelectric $PbZr_{0.2}Ti_{0.8}O_3$

Jingfeng Song[1], Zhiyong Xiao[1], Bo Chen[2], Spencer Prockish[1], Xuegang Chen[1], Jinsong Huang[2,3] and Xia Hong[1,3*]

[1]Department of Physics and Astronomy, University of Nebraska-Lincoln, Nebraska 68503-0299

[2]Department of Mechanical and Materials Engineering, University of Nebraska-Lincoln, Nebraska 68588-0526

[3]Nebraska Center for Materials and Nanoscience, University of Nebraska-Lincoln, NE 68588-0298

* Email: xia.hong@unl.edu

## Abstract

We report a more than 10-fold enhancement of the piezoelectric coefficient $d_{33}$ of polycrystalline $CH_3NH_3PbI_3$ ($MAPbI_3$) films when interfacing them with ferroelectric $PbZr_{0.2}Ti_{0.8}O_3$ (PZT). Piezo-response force microscopy (PFM) studies reveal $d_{33}^{MAPbI_3}$ values of ~0.3 pm/V for $MAPbI_3$ deposited on Au and ITO surfaces, with small phase angles fluctuating at length scales smaller than the grain size. In sharp contrast, on samples prepared on epitaxial PZT films, we observe large scale polar domains exhibiting clear, close to ±90° PFM phase angles, pointing to polar axes along the film normal. By separating the contributions from the $MAPbI_3$ and PZT layers, we extract a significantly enhanced $d_{33}^{MAPbI_3}$ value of ~4 pm/V, which is attributed to the long-range dipole-dipole interaction facilitated domain nucleation. We also discuss the effect of the interfacial screening layer on the preferred polar direction.

The organolead trihalide perovskites such as $CH_3NH_3PbI_3$ ($MAPbI_3$) have gained significant research interests in recent years due to its high power conversion efficiency [1-3], strong solar absorption [1, 2], and long charge diffusion lengths [4-6], making it a promising material choice for developing high performance photovoltaic applications. It has been shown theoretically that these materials are polar [7, 8], which is corroborated by the experimental observations of the piezoelectricity [9-13] and ferroelasticity [14] in both $MAPbI_3$ polycrystalline films and single crystals. Photoinduced enhancement in the piezoelectric coefficient $d_{33}$ [8, 9] and photostriction effect [15] in these materials suggest a strong connection between the polar nature and their optical response. A highly debated proposal to account for the high power conversion efficiency involves the existence of randomly oriented ferroelectric domains, where the internal field within the domain wall (DW) can promote the photocarrier separation and reduce recombination [3, 7, 16], similar to the mechanism driving the high photovoltaic voltage in the multiferroic oxide $BiFeO_3$ [17, 18]. While extensive research has been carried out to probe ferroelectricity in $MAPbI_3$ [9, 13, 19, 20], the polarization switching mechanism and its role in the power conversion process are yet to be unambiguously identified. The ability to control and manipulate the polar properties in these materials thus has important implications in gaining enhanced understanding of its photovoltaic response.

In addition, it is also of high technological interest to explore the piezoelectric properties of the hybrid perovskites for applications such as piezoelectric generators or energy harvesting devices [12, 13]. For $MAPbI_3$ single crystals, a $d_{33}$ value of 2.7 pm/V has been reported from direct optical measurement on large single crystals [11], close to the value for quartz (~2.0 pm/V) [21], but significantly smaller than the values for other widely used piezoelectric materials such as ZnO (12 pm/V) [22] and $Pb(Zr,Ti)O_3$ (up to 400 pm/V) [23]. Local $d_{33}$ value of 6 pm/V was obtained via piezo-response force microscopy (PFM) in polycrystalline thin films [9], while the average response of the sample over large scale can be compromised due to the random orientated polar axes in different grains. To date, no effective material strategy has been developed to engineer the polar axis in this material for optimized piezoelectric response.

In this work, we report a more than 10-fold enhancement in the piezoelectric coefficient of polycrystalline MAPbI$_3$ thin films when interfacing with an epitaxial ferroelectric PbZr$_{0.2}$Ti$_{0.8}$O$_3$ (PZT) layer. PFM studies show that MAPbI$_3$ films prepared on Au and indium tin oxide (ITO) surfaces exhibit $d_{33}^{MAPbI_3}$ values of ~0.3 pm/V, with small PFM phase angles fluctuating at length scales smaller than the film grain size. For samples deposited on PZT, in sharp contrast, we observed large scale domain structures with clear, close to ±90° PFM phase angles, suggesting the polar axes are along the film normal. Quantitative PFM amplitude studies reveal both constructive and destructive piezo-responses between the MAPbI$_3$ and PZT layers. By separating their individual contributions using finite element analysis, we extracted a $d_{33}^{MAPbI_3}$ value of ~4 pm/V for both polar states. The significantly enhanced domain size and piezoelectric response are attributed to the interfacial dipole-dipole interaction with the ferroelectric polarization of PZT, which promotes the alignment of the polar axis of MAPbI$_3$ at the domain nucleation stage. We also propose that the preferred polar direction depends on the strength of interfacial screening.

We worked with PZT, Au and ITO thin films as the base layer for MAPbI$_3$ deposition. Epitaxial heterostructures composed of 50 nm PZT and 6 nm La$_{0.67}$Sr$_{0.33}$MnO$_3$ (LSMO) were deposited on (001) SrTiO$_3$ (STO) substrates (Fig. 1(a)), where the conductive LSMO layer serves as the bottom electrode. The PZT films have smooth surfaces with root mean square (RMS) roughness of ~5 Å, with the as-grown polarization along film normal in the uniform up orientation ($P_{up}$) [24]. We deposited 20-300 nm polycrystalline MAPbI$_3$ films on different base layers via spin coating followed by thermal annealing at 100°C, with typical grain sizes of 200-400 nm. The AFM and PFM studies were carried out using a Bruker Multimode 8 AFM system in $N_2$ atmosphere to minimize moisture-induced sample degradation [25]. The PFM measurements were taken in contact mode, with an AC voltage $V_{bias}$ applied to the conductive AFM tip and the bottom electrode grounded. We conducted PFM imaging close to the resonant frequency of the cantilever (300±20 kHz) with $V_{bias}$ = 0.15 V and acquired the $d_{33}$ values well below the resonance frequency (50 kHz). There is less than 20% change in the extracted $d_{33}$ between 20-50 kHz. The high frequency employed help eliminate the artifacts due to ionic electromigration within $MAPbI_3$ [26]. The details of sample preparation and characterization can be found in Refs. [24, 27, 28] and the Supplementary Materials [29].

Figure 1(b) shows the AFM topography image of a 300 nm $MAPbI_3$ thin film deposited on PZT. This area has a relatively smooth morphology, with the RMS surface roughness of ~20 nm. The majority of the area shows a mosaic-like PFM phase landscape (Fig. 1(c)), with the majority of area having phase angles around +90°. As the underlying PZT film is uniformly polarized in the $P_{up}$ state with homogeneous PFM response [29], such phase variation can only be intrinsic to the $MAPbI_3$ sample, reflecting the existence of domain structures with different polar axis orientations. Similar spatial inhomogeneity in the PFM phase response was also observed on films prepared on Au and ITO surfaces, although the phase angle is much smaller and fluctuating around 0° [29]. It is worth noting that the typical size for regions with uniform phase and amplitude (Fig. 1(d)) responses is smaller than the grain size (Fig. 1(b)). This points to the existence of polar domains within a grain, suggesting that the DW energy in $MAPbI_3$ is relatively low [30]. The high phase angle also suggests that the polar axis for samples on PZT is close to the surface normal.

More interestingly, embedded in a background that is dominated by positive phase response, there is a region close to the bottom of the image that shows a clear phase signal of around -90° (highlighted in Fig. 1(c)). This region corresponds to a relatively smooth surface morphology (Fig. 1(b)), thus excluding artifacts in PFM imaging due to sharp step edges or sample pinholes. As shown in Fig. 1(d), the corresponding PFM amplitude is significantly lower than the regions with positive phases (Fig. 1(d)). Such large scale domains exhibiting close to ±90° phase angles are absent in samples on Au and ITO, suggesting that it is the underlying PZT layer that promotes the nucleation and growth of domains with the polar axis close to the surface normal.

As $V_{bias}$ is applied across the MAPbI$_3$/PZT heterostructure, the PZT layer would also contribute to the overall PFM response. It is thus important to identify the relative alignment between the polar axes of MAPbI$_3$ and PZT, which may result in either constructive or destructive PFM responses. To gain precise knowledge of the polar direction of MAPbI$_3$, we focused the study on thinner samples (20-60 nm) close to the film boundary, so we can measure the regions of bare PZT at the same time. Figure 2(a) shows the topography image of a sample with thickness of 25±5 nm, where we observed regions with PFM phase structures (Fig. 2(b)) that cannot be fully correlated with the sample morphology. Besides mosaic phase variation, similar to those observed on the thick films, we also observed large scale domains exhibiting clean, uniform signal close to ±90° with the domain size extending over several grains. The bare PZT region, on the other hand, shows homogenous phase angle of -90°, corresponding to the $P_{up}$ state. The majority of the large domains in MAPbI$_3$ shows PFM responses close to +90°, which is out-of-phase with the signal from the PZT film. There are, however, several isolated regions with phase angle of about -90°, in-phase with that of PZT.

We then carried out quantitative PFM amplitude analysis on areas with different phase responses to extract the corresponding $d_{33}$ values. In Fig. 2(a), we focused on two spots showing +90° (blue cross) and -90° (red circle) phase responses (Fig. 2(b)). As both spots reside at the central area of a grain, we can rule out artifacts due to grain boundary induced ion migration, pinholes or sharp height variation [31, 32]. On each site, we then measured the PFM amplitude response $A$ with $V_{bias}$ ramped from 0 to 3 V at 0.1 V/s (Fig. 2(d)). For both spots, $A(V_{bias})$ exhibits a quasi-linear behavior, while the signal on the region in-phase with PZT (-90°) is consistently higher than that on the region out-of-phase with PZT (+90°). At $V_{bias}$ = 3 V, the amplitude responses of these two spots reach 4 mV and 2 mV, respectively. For bare PZT (black diamond in Fig. 2(a)), the amplitude reaches 4 mV at $V_{bias}$ = 1 V. We also showed the data taken on a 25 nm film prepared on Au for comparison, whose amplitude signal only reaches ~0.28 mV at $V_{bias}$ = 3 V, similar to the sample on ITO [29].

To calculate $d_{33}$, we converted the amplitude $A$ into the sample displacement $u_{tot} = AS/I$, where $S$ =50 pm/V is the tip sensitivity and $I$ =16 is the vertical gain. We extracted $d_{33} = \partial u_{tot}/\partial V_{bias}$ from the initial slope of $A(V_{bias})$, or the linear component of the PFM response [9, 11, 13]. For bare PZT, we obtained a $d_{33}^{PZT}$ of 19±2 pm/V, comparable with reported value obtained for PZT films with similar composition using optical method [33]. A super-linear dependence emerges at high $V_{bias}$ voltages, which can be attributed to a quadratic contribution of the ferroelastic effect [34]. For $MAPbI_3$ film on Au and ITO, the $d_{33}$ value is about 0.3 pm/V, almost two orders of magnitude lower than that of PZT [29].

For the $MAPbI_3$/PZT heterostructure, we need to first separate the piezo-responses from the $MAPbI_3$ and PZT layers. We performed the finite element analysis to calculate the voltage drops across the $MAPbI_3$ and PZT layers [35]. Figure 3(a) shows the simulated potential distribution through a heterostructure with 20 nm $MAPbI_3$ and 50 nm PZT at $V_{bias}$ = 1 V, assuming a point contact with the AFM tip and a global ground provided by the LSMO layer. For modeling, we used dielectric constants of 100 for PZT [36] and 32 for $MAPbI_3$ [6] based on our previous studies. Using a higher dielectric constant of 50 for $MAPbI_3$ only changes the result by less than 4%. The potential profile along the central dashed line is shown in Fig. 3(b). For our sample geometry, the majority of the bias voltage is always applied through the MAPbI$_3$ layer, with the fraction of voltage ($v_{MA} = V_{MAPbI_3}/V_{bias}$) changing from 87% to 93% as the MAPbI$_3$ film thickness increases from 20 nm to 60 nm (Fig. 3(c)).

Once we determined the voltage fractions across the MAPbI$_3$ and PZT ($v_{PZT} = V_{PZT}/V_{bias}$) layers, we calculated the $d_{33}$ value of MAPbI$_3$ using:

$$\frac{\partial u_{tot}}{\partial V_{bias}} = d_{33}^{MAPbI_3} v_{MA} \pm d_{33}^{PZT} v_{PZT} . \tag{1}$$

Here we $d_{33}^{PZT}$ =19 pm/V for PZT, "+" for constructive (in-phase) and "-" for destructive (out-of-phase) piezoelectric responses between these two layers. For the sample shown in Fig. 2, we extracted $d_{33}^{MAPbI_3}$ = 3.7±0.1 pm/V for the region that is in-phase with the piezo-response of bare PZT and 4.2±0.2 pm/V for the out-of-phase region. We have conducted the PFM ramp studies on three MAPBI$_3$ samples, with thickness varying from 20 to 60 nm. On all samples, we observed similar large domain structures with close to ±90° PFM phase angles, suggesting a predominant existence of polar axis along the film normal [29]. The piezoelectric coefficient was found to be about 4 pm/V for both polarization directions, with no appreciable dependence on the film thickness (Fig. 4(a)). The films on Au and ITO, on the other hand, reveal consistently a low $d_{33}$ value of about 0.3 pm/V. This value is about one order of magnitude lower than that obtained on bulk single crystal samples [11], suggesting a random, isotropic distribution of the polar axes, which gives rise to a small out-of-plane component of the piezoelectric response. This picture is supported by the small, highly fluctuating phase angle observed in these samples [29].

The distinct difference in the amplitude and phase angle of the piezoelectric response in samples prepared on different substrates suggests that a ferroelectric base layer favors out-of-plane polar axis in MAPBI$_3$ films. The evolution from large scale domains (Figs. 2(b)-(d)) to a mosaic domain structure (Figs. 1(c)-(d)) with increasing MAPBI$_3$ film thickness also points to the interfacial nature of the interaction. A possible mechanism for the enhanced piezoelectric response in MAPbI$_3$ thin films is that the interfacial dipole-dipole interaction with the strong polarization field of PZT promotes the alignment of the polar axis in the initial domain nucleation stage. Once the domains with $P_{up}$ and $P_{down}$ states are seeded, it is energetically unfavorable for the growth of domains with other polar axis, as it requires the formation of charged DW. The long-range dipole interaction is known to be an important energy scale in ferroelectric materials that competes with the elastic energy in determining the DW energy and domain structures in epitaxial $PbZr_{0.2}Ti_{0.8}O_3$ thin films [37] and layered ferroelectric copolymer poly(vinylidene-fluoride-trifluorethylene) [P(VDF-TrFE)] [30, 38]. This effect is especially pronounced in P(VDF-TrFE), where the inter-layer bonding is van der Waals type and weak, and the dipole-dipole interaction is the driving force to align the out-of-plane polarization between different monolayers and suppress the DW roughness [38]. Enhanced local piezoelectric response has been achieved in P(VDF-TrFE) thin films by optimizing the local dipole alignment through precise microstructure control [39], consistent with the observation in the present study. Similar scenario has also been use to explain the enhanced open voltage and photovoltaic efficiency in $BiFeO_3$ when interfaced with $MAPbI_3$ [18].

The dipole-dipole interaction alone, however, cannot explain why a large fraction of the domains have the polar axis anti-aligned with the polarization of PZT, which seems to lead to an increased, rather than reduced electrostatic energy. What has not been taken into account is the existence of an interfacial charge screening layer from the ambient, such as molecular or dissociated water or other charged adsorbates, prior to $MAPBI_3$ deposition, as $MAPbI_3$ was crystallized well below the ferroelectric Curie temperature of PZT [37]. In previous studies of ferroelectric oxide/graphene hybrid devices, it has been shown that the switching hysteresis in graphene can be either in-phase or out-of-phase with the ferroelectric switching, depending on whether the interfacial charge layer provides under-screening or over-screening to the ferroelectric polarization field [40]. We thus propose that it is the screening strength of this interfacial charge layer that determines the preferred polar axis of MAPbI$_3$. When the interfacial charge layer is not sufficient to completely screen PZT's polarization, the residue polarization would facilitate the nucleation of the MAPbI$_3$ domains with polar axis aligned with PZT's polarization (Fig. 4(b)). In the case of over-screening, on the other hand, the MAPbI$_3$ layer would sense a secondary electric field generated by the interfacial charge, which actually favors domain nucleation with polar axis anti-aligned with PZT's polarization (Fig. 4(c)). At room temperature, we expect the polarization of PZT is close to be completely screened. Upon the crystallization of MAPbI$_3$ films at 100°C, the polarization of PZT would decrease due to the pyroelectric effect [40]. Assuming the interfacial charge density cannot change promptly, the interface would be dominated by the over-screening condition, which naturally explains why the favored polar axis is anti-aligned with PZT's polarization. This scenario also suggests that it is possible to engineer the polar axis of polycrystalline MAPbI$_3$ thin films in a uniform state by controlling the surface screening condition of a ferroelectric base layer.

In summary, working with MAPbI$_3$/PZT heterostructures, we have shown that the presence of an interface polarization can promote the alignment of the polar axis of grains in polycrystalline MAPbI$_3$ thin films, leading to a more than 10-fold enhancement in their piezoelectric response. Our study provides an effective material strategy to control the polar properties of organolead trihalide perovskite thin films. It also paves the path for combining the high efficiency photovoltaic effect in the hybrid perovskites with the rich functionality of complex oxides for developing integrated electronic, mechanical, and optoelectronic applications.

J.S. and Z.X. contributed equally to this work. The authors would like to thank Angus Kingon and Nitin P. Padture for insightful discussions and Dong Wang for experimental assistance. This work was supported by the NSF Nebraska Materials Research Science and Engineering Center (MRSEC) Grant No. DMR-1420645 (sample preparation), NSF CAREER Grant No. DMR-1148783 (oxide thin film characterization), and NSF Grant No. OIA-1538893 (scanning probe studies and manuscript preparation). S.P. acknowledges support from Nebraska Public Power District through the Nebraska Center for Energy Sciences Research. The research was performed in part in the Nebraska Nanoscale Facility: National Nanotechnology Coordinated Infrastructure and the Nebraska Center for Materials and Nanoscience, which are supported by the National Science Foundation under Award ECCS: 1542182, and the Nebraska Research Initiative.

# Figure 1

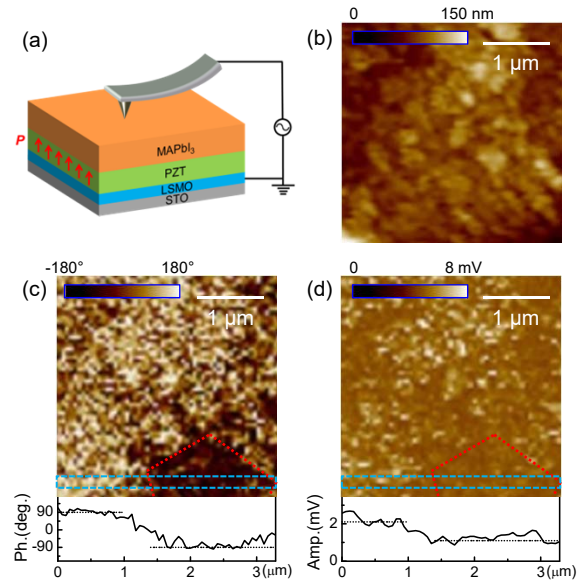

**FIG. 1.** (**a**) Schematic view of MAPbI$_3$ on PZT. (**b**) AFM topography, (**c**) PFM amplitude and (**d**) phase images taken on a 300 nm MAPbI$_3$ film on PZT. The lower panels show the signal line profiles averaged over the region enclosed in the dashed boxes. The dotted lines serve as a guide to the eye.

# Figure 2

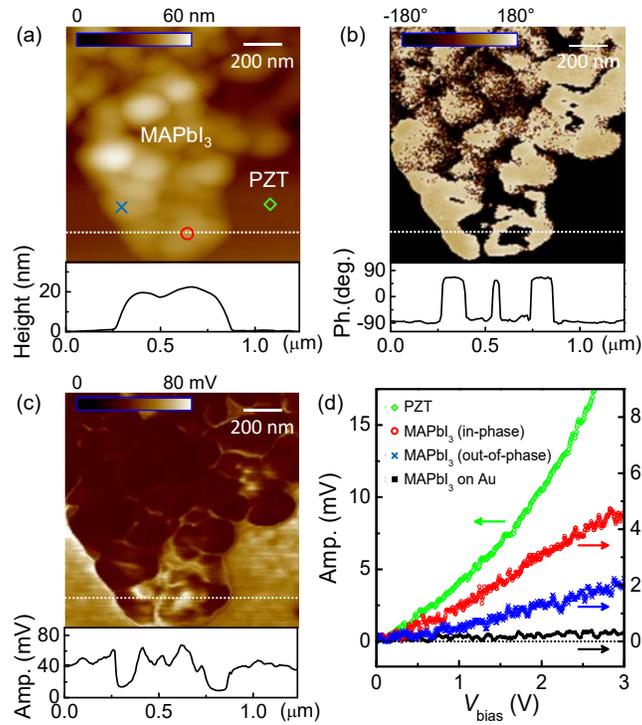

**FIG. 2.** (**a**) AFM topography, (**b**) PFM phase and (**c**) amplitude images taken on a 25±5 nm MAPbI$_3$ film on PZT. The lower panels show the signal profiles along the dotted lines. (**d**) PFM amplitude vs. $V_{bias}$ measured on spots marked in (**a**), and the data taken on a 25 nm MAPbI$_3$ on Au.

Figure 3

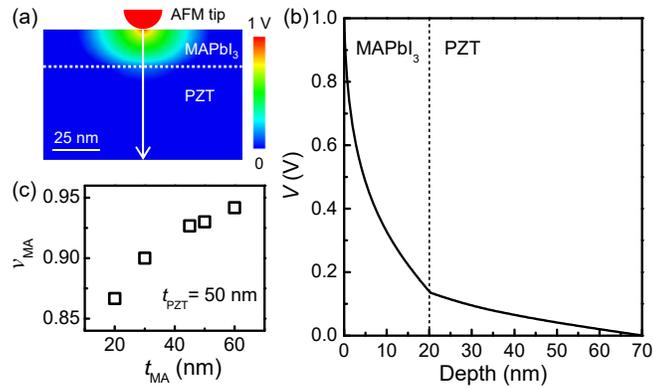

**FIG. 3.** (**a**) Simulated potential distribution in a 20 nm MAPbI$_3$/50 nm PZT heterostructure with $V_{bias}$ = 1 V. (**b**) The voltage drop along the central solid line in (**a**). (**c**) The fractional voltage across MAPbI$_3$ as a function of MAPbI$_3$ film thickness.

Figure 4

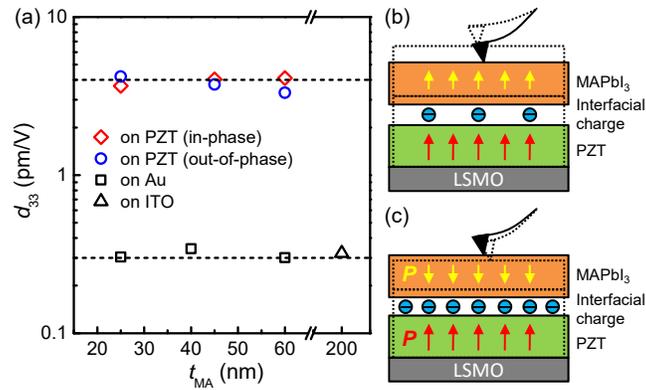

**FIG. 4.** (**a**) $d_{33}$ vs. MAPbI$_3$ film thickness for samples on Au, ITO and PZT. The dashed lines serve as a guide to the eye. (**b**) Schematic view for constructive piezoelectric response between MAPbI$_3$ and PZT with an under-screening interfacial charge layer. (**c**) Schematic view for destructive piezoelectric response between MAPbI$_3$ and PZT with an over-screening interfacial charge layer.